\documentclass[twocolumn,superscriptaddress,showpacs,nofootinbib,notitlepage]{aastex631}
\usepackage{graphicx,subfigure,amsmath,amsfonts,amssymb,multirow,mathrsfs,ulem,threeparttable}
\usepackage{hyperref}
\usepackage{epstopdf}
\begin{document}
\title{Digging into the ultraviolet luminosity functions of galaxies at high redshifts: galaxies evolution, reionization, and cosmological parameters}

\correspondingauthor{Yi-Zhong Fan}
\email{yzfan@pmo.ac.cn}

\author[0000-0003-1215-6443]{Yi-Ying Wang}
\affiliation{Key Laboratory of Dark Matter and Space Astronomy, Purple Mountain Observatory, Chinese Academy of Sciences, Nanjing 210033, People's Republic of China}
\affiliation{School of Astronomy and Space Science, University of Science and Technology of China, Hefei, Anhui 230026, People's Republic of China}

\author[0000-0003-4631-1915]{Lei Lei}
\affiliation{Key Laboratory of Dark Matter and Space Astronomy, Purple Mountain Observatory, Chinese Academy of Sciences, Nanjing 210033, People's Republic of China}
\affiliation{School of Astronomy and Space Science, University of Science and Technology of China, Hefei, Anhui 230026, People's Republic of China}

\author[0000-0001-9120-7733]{Shao-Peng Tang}
\affiliation{Key Laboratory of Dark Matter and Space Astronomy, Purple Mountain Observatory, Chinese Academy of Sciences, Nanjing 210033, People's Republic of China}

\author[0000-0002-4538-8526]{Guan-Wen Yuan}
\affiliation{Department of Physics, University of Trento, Via Sommarive 14, 38123 Povo (TN), Italy}
\affiliation{Trento Institute for Fundamental Physics and Applications (TIFPA)-INFN, Via Sommarive 14, 38123 Povo (TN), Italy}

\author[0000-0002-8966-6911]{Yi-Zhong Fan}
\affiliation{Key Laboratory of Dark Matter and Space Astronomy, Purple Mountain Observatory, Chinese Academy of Sciences, Nanjing 210033, People's Republic of China}
\affiliation{School of Astronomy and Space Science, University of Science and Technology of China, Hefei, Anhui 230026, People's Republic of China}

\newcommand{\ud}{\mathrm{d}}
\begin{abstract}
Thanks to the successful performance of the James Webb Space Telescope, our understanding of the epoch of reionization of the Universe has been advanced. The ultraviolet luminosity functions (UV LFs) of galaxies span a wide range of redshift, not only revealing the connection between galaxies and dark matter (DM) halos but also providing the information during reionization. In this work, we develop a model connecting galaxy counts and apparent magnitude based on UV LFs, which incorporates redshift-dependent star formation efficiency (SFE) and corrections for dust attenuation. 
By synthesizing some observations across the redshift range $4\le z \le 10$ from various galaxy surveys, we discern the evolving SFE with increasing redshift and DM halo mass through model fitting. Subsequent analyses indicate that the Thomson scattering optical depth was $\tau_{\rm e} = 0.054^{+0.001}_{-0.003}$ and the epoch of reionization started (ended) at $z=18.8^{+7.2}_{-6.0}$ ($z=5.3^{+0.8}_{-1.0}$) which is insensitive to the choice of the truncated magnitude of the UV LFs. Incorporating additional dataset and some reasonable constraints, the amplitude of matter perturbation is found to be $\sigma_8=0.80\pm0.05$, which is consistent with the standard $\Lambda$CDM model. Future galaxy surveys and the dynamical simulations of galaxy evolution will break the degeneracy between SFE and cosmological parameters, improving the accuracy and the precision of the UV LF model further.
\end{abstract}

\section{Introduction}\label{sec:1}
Following the formation of the first stars, the Universe emerged from its dark ages as the neutral hydrogen gas gradually became ionized under the illumination of the first lights. Present observations have constrained the bulk of reionization to be completed by $z\sim6$ \citep{2006AJ....132..117F}, but the beginning of reionization shows less well-constraint and depends sensitively on the nature of the ionizing sources. During the epoch of reionization, there has been debate regarding the primary energy sources responsible for rendering the universe transparent. \citet{2015ApJ...802L..19R} analyzed the Thomson scattering optical depth $\tau$ from Planck and the abundance and luminosity distribution of distant galaxies from Hubble Space Telescope imaging. Their results strengthened the assumption that most of the ionizing photons responsible arose from high-redshift star-forming galaxies. Subsequently, \citet{2019ApJ...879...36F} demonstrated that galaxies become more efficient producers of ionizing photons at higher redshifts and fainter magnitudes. The contribution of AGNs represented subdominant. A recent study \citep{2022NatAs...6..850J} found that the contribution of quasars was negligible, providing less than $7\%$ of the total required photons. Based on the ultra-deep James Webb Space Telescope (JWST) imaging, \citet{2024Natur.626..975A} analyzed eight ultra-faint galaxies and demonstrated that the majority of the necessary photons originated from dwarf galaxies, highlighting the significant connection between reionization and galaxies. However, at $z \gtrsim 9$, the observations of JWST suggest a tension between the current data and the cosmic microwave background, since these high-redshift galaxies with $M_{\rm UV} <-15 \, \rm mag$ not only can drive reionization but would end it too early \citep{2024arXiv240407250M}. Besides, researches on the evolution of the neutral hydrogen fraction ($x_{\rm HI}$) across different redshifts yielded various results \citep{2018ApJ...854...73I, 2019ApJ...879...36F, 2020ApJ...892..109N}, accompanied by notable uncertainties and discrepancies.
For instance, \citet{2024ApJ...973....8H} imposed a constraint of $x_{\rm HI}>0.9$ at $z=10.17$ whereas \citet{2023ApJ...949L..40B} derived $x_{\rm HI}<0.88$ at $z=10.6$. The UV luminosity functions (UV LFs) of galaxies span a wide range of redshift from cosmic dawn to cosmic noon. Currently, the observations of UV LFs have extended up to $z\sim15-16$ \citep{2023MNRAS.523.1036B,2023ApJS..265....5H,2024MNRAS.533.3222D}, offering opportunities to explore the epoch of reionization.

Since galaxies were formed inside the dark matter (DM) halos, their abundance intricately tied to the distribution of DM halos \citep{2003ApJ...593....1B}. The UV LFs represent the number density of galaxies at specific luminosity, serving as an observable metric that links galaxies with their DM halos. This connection unveils the underlying physical mechanisms governing galaxy formation and evolution, particularly the star formation efficiency (SFE), which quantifies the effectiveness of converting gas into stars. Based on the assumption that the most massive galaxies inhabit the most massive DM halos, the SFE of central galaxies can be derived by the abundance matching between the DM mass function and UV LFs \citep{2003ApJ...585L.117B,2009ApJ...695..900Y}. As DM halo masses increase, the profile of SFE reflects the influence from diversified feedback processes, including AGN feedback \citep{2006MNRAS.365...11C}, supernovae feedback \citep{2012MNRAS.421.3522H}, dynamical friction effect and others \citep{2006MNRAS.368....2D}. Only in the local universe around $z\sim0$, the evolution of the SFE with increasing DM halo mass has almost been ascertained (see \citet{2018ARA&A..56..435W} for a review).
However, it remains unclear whether the SFE evolves with redshift, and if it does, the nature of its evolutionary trajectory is yet to be unknown. 
Nevertheless, by establishing connections between DM halo mass and galaxy luminosity (or mass), UV LFs provide a promising avenue to extend SFE investigations to higher redshift ranges \citep{2018MNRAS.477.1822M, 2022ApJS..259...20H, 2023ApJ...954L..48W}.

Besides exploring the epoch of reionization and the SFE, UV LFs provide additional capabilities for cosmological constraints \citep{2003MNRAS.339.1057Y,2017MNRAS.470.2100K}. At high redshift, UV LFs possess the ability to extract the information within Mpc scales. For instance, \citet{2022ApJ...928L..20S} measured the matter power spectrum $p(k)$ at wavenumbers of $0.5 \, {\rm Mpc^{-1}} < k < 10 \, {\rm Mpc^{-1}}$ to roughly $30\%$ precision by UV LFs at the range of $4 \le z \le 10$. Such constraints can be derived from the cumulative stellar mass distributions \citep{2023MNRAS.526L..63P}, also. Given that the number density of DM halos is sensitive to the amplitude of mass fluctuations $\sigma_8$, UV LFs can be used to constrain $\sigma_8$ at high redshift, alleviating the $\sigma_8$ tension potentially \citep{2022PhRvD.105d3518S} (a software named GALLUMI). Furthermore, UV LFs provide a way to examine theoretical scenarios beyond the $\Lambda$CDM model, such as the warm DM, the ultralight axion DM and the effect of primordial non-Gaussianities \citep{2015MNRAS.446.3235C,2017PhRvD..95h3512C,2024arXiv240411071W}.

In this work, we analyze the UV LFs data with a universal model to depict the evolution of galaxies and the process of reionization. To mitigate the biases arising from differing cosmological framework\footnote{Different observations may based on different cosmological assumptions. For instance, the usual choices include $H_0=70 \, \rm km \, s^{-1} \, Mpc^{-1}$, $\rm \Omega_{m} =0.3$ and $H_0=67.6 \, \rm km \, s^{-1} \, Mpc^{-1}$, $\rm \Omega_{m} =0.307$.}, we rescale the observations and the UV LFs model to the galaxy counts and apparent magnitude. Differing from our previous research \citep{2023ApJ...954L..48W}, the function of SFE in this study is redshift-dependent and demonstrates a distinct evolution with increasing redshift.
In that case, the feedback strength that suppresses the birth of stellar displays a more continuous variation. The derived UV luminosity density is tightly constrained within a narrow range and is well consistent with findings from other studies. Additionally, by incorporating supplementary observations, we enhance constraints on the entire evolution of $x_{\rm HI}$, the beginning redshift of reionization and the Thomson scattering optical depth $\tau_{\rm e}$. We also estimate the values of $H_0$ and $\rm \Omega_{m}$ using an additional dataset, as these two cosmological parameters can not be effectively constrained by the UV LFs observations alone. 
Furthermore, we constrain $\sigma_8$ within a reasonable range based on plausible assumptions. 

\section{Method}\label{sec:2}
Building on previous works \citep[e.g.,][]{2022ApJ...938L..10I,2023MNRAS.525.3254S,2023ApJ...953L...4P}, we refine the UV LF model $\Phi(M_{\rm UV})$ to elucidate the relationship between the distribution of DM halos and the UV LFs observations. Our analysis spans redshifts ranging from 4 to 10, adopting the DM mass function $\phi(M_{\rm h})$ characterized by the ``Sheth-Mo-Tormen" function \citep{2001MNRAS.323....1S}. The calculations of the transfer function within $\phi(M_{\rm h})$ are conducted using the {\tt CAMB} software \citep{2000ApJ...538..473L}, which accommodates all parameter spaces of the Friedmann-Robertson-Walker models. The mathematical representation of the DM mass function is readily available through the Python package {\tt HMFcal} \citep{2013A&C.....3...23M}, which supports computations across multiple cosmological frameworks. Subsequently, the DM mass function is transformed into $\Phi(M_{\rm UV})$ using
\begin{equation}\label{eq:1}
\Phi(M_{\rm UV}) = \phi(M_{\rm h})\bigg| \frac{{\rm d}M_{\rm h}}{{\rm d}M_{\rm UV}} \bigg|,
\end{equation}
where $\big| \frac{{\rm d}M_{\rm h}}{{\rm d}M_{\rm UV}} \big|$ is the Jacobian determinant. In the AB magnitude system, $M_{\rm UV}$ denotes the intrinsic ultraviolet magnitude \citep{1983ApJ...266..713O}, which is related to the UV luminosity $L_{\rm UV}$ according to the equation
\begin{equation}\label{eq:2}
M_{\rm UV}=-2.5\log_{10} \bigg( \frac{L_{\rm UV}}{\rm erg \, s^{-1}} \bigg)+51.63.
\end{equation}
The UV luminosity can be straightforwardly derived from the star formation rate (SFR) \citep{2014ARA&A..52..415M}. Employing the Salpeter initial mass function (IMF) \citep{1955ApJ...121..161S} at a wavelength of $\lambda_0=1500\rm \AA$, the specific UV luminosity is expressed as
\begin{equation}
L_{\rm UV} = \frac{1}{1.15}\times10^{28} \times \bigg[ \rm \frac{erg \, s^{-1}\,Hz^{-1}}{yr^{-1}\,M_{\odot}} \bigg] \times{\rm SFR}.
\end{equation}
Specifically, the SFR is contingent upon the total baryonic inflow rate $\dot{M}_{\rm b}$ into the DM halo and the star formation efficiency (SFE) $\epsilon$. Consequently, the SFR can be formulated as ${\rm SFR}=\epsilon \dot{M}_b$, where $\dot{M}_{\rm b}$ is defined as $\dot{M}_{\rm h} f_{\rm b}$, with $f_{\rm b}$ being the baryonic fraction $\Omega_b/\Omega_m=0.156$ \citep{2020A&A...641A...6P}.

Different from our previous work \citep{2023ApJ...954L..48W}, which focused on calculating the DM mass accretion rate (MAR) within the $\Lambda$CDM framework, this study employs a more comprehensive accretion model to facilitate the analysis of cosmological parameters. We utilize an alternative analytical method, the extended Press-Schechter (EPS) theory \citep{1991ApJ...379..440B, 1991MNRAS.248..332B, 1993MNRAS.262..627L}, to characterize the DM accretion histories. \citet{2024MNRAS.52711740L} conducted a comparison between two analytical models based on the EPS theory and three empirical models derived from cosmological simulations. Their findings confirmed the accuracy of the differential equation method for describing mass growth ($\dot{M}_{\rm h}={\rm d}M_{\rm EPS}/{\rm d}z$) \citep{2015MNRAS.450.1514C}. Therefore, we adopt their equations (A5-B7) for calculating mass accretion, incorporating adjustable cosmological parameters.

In the redshift range of $4\le z \le 10$, we have collected observations of UV LFs from various surveys conducted by multiple telescopes. These include JWST (SMACS0723, GLASS, CEERS, COSMOS, NGDEEP, HUDF, CANUCS and JADES; \citet{2023MNRAS.523.1036B, 2023MNRAS.518.6011D, 2024ApJ...966...74W, 2024ApJ...965..169A, 2023ApJ...951L...1P}), the Hubble Space Telescope (HST) (the Ultra-Deep Field, the GOODS fields, the Hubble Frontier Fields parallel fields, and all five CANDELS fields; \citet{2013MNRAS.432.2696M,2015ApJ...810...71F,2018ApJ...855..105O,2021AJ....162...47B,2022ApJ...940...55B}) , the Subaru/Hyper Suprime-Cam survey and CFHT Large Area U-band Survey \citep{2022ApJS..259...20H,2023ApJ...946L..35M,2023MNRAS.523..327A}, the Visible and Infrared Survey Telescope for Astronomy (VISTA) (COSMOS, XMM-LSS, VIDEO, and Extended Chandra Deep Field South (E-CDFS) field; \citet{2020MNRAS.493.2059B, 2023MNRAS.523..327A}), and the UK Infrared Telescope (UKIRT) (UKIDSS UDS field; \citet{2020MNRAS.493.2059B}).

Prior to analysis, overlaps in magnitudes, fields, and redshift bins were eliminated to prevent duplication. At first, for various telescope projects, we artificially collect observation catalogues that include high redshift ($z\ge4$) galaxies as many as possible. As listed above, most of these catalogues exhibit independence because their survey fields are completely irrelevant. Subsequently, a small fraction of the observations are removed to eliminate the effect of the non-negligible overlap. For example, at $z=9$, \citet{2023ApJ...951L...1P} derived the UV LFs from deep NIRCam observations taken in parallel with the MIRI Deep Survey of the HUDF, which has been incorporated into \citet{2021AJ....162...47B} base on HST. In that case, the observations in \citet{2023ApJ...951L...1P} with $M_{\rm UV}<-17.92$ (the faintest observations in  \citet{2021AJ....162...47B}) are excluded. After such screening processes, all of the UV LF data we used are mutually independent. 

Due to the fluctuations in the large-scale dark matter density field, cosmic variance generates the intrinsic scatter in the number density of galaxies and then introduces unavoidable errors into the UV LFs. Especially at high redshift and small survey areas, cosmic variance can be very significant. To address this, we verify whether such uncertainties have been considered in all of the UV LF observations. For the dataset that did not account for cosmic variance, we add the additional uncertainties \footnote{Given the parameter limitation of {\tt GALCV}, we conservatively assume $z=5$ to calculate the cosmic variance for the observations at $z=4$. For wider survey area ($> 31640 \, \rm arcmin^2$), cosmic variance is estimated using {\tt LINCV}. Both of the methods are available at \url{https://github.com/adamtrapp/galcv}.} using the public Python package {\tt GALCV} \citep{2020MNRAS.499.2401T}.  For a clearer comparison across different magnitudes, redshifts, and field areas, we present the data from \citet{2021AJ....162...47B} along with their corresponding cosmic variance in \autoref{Tab:4}. All UV LF data used in this analysis can be accessed via \href{https://github.com/wangyy19/UVLF_datasets}{GitHub} \footnote{\url{https://github.com/wangyy19/UVLF_datasets}}. 

Nevertheless, the UV flux emitted by galaxies is invariably absorbed by interstellar dust, especially in massive galaxies and at lower redshifts. Due to this dust extinction, such observations cannot be directly integrated into the UV LFs model. Moreover, it is necessary to harmonize all observations within a consistent cosmological framework, as differing constructions of the UV LFs rely on varied cosmological parameters, specifically matter density ($\Omega_{\rm m}$) and the Hubble constant ($H_0$).

\subsection{Calibrating the observations to intrinsic magnitude}

At first, we address the issue of dust extinction. For a spectrum modeled as $f_{\rm \lambda} \sim {\rm \lambda}^{\beta}$, the UV-continuum slope $\beta$ and its magnitude are linearly related as $\langle \beta \rangle = ({\rm d}\beta/{\rm d}M_{\rm UV})[M_{\rm UV}-M_0]+\beta_0$ \citep{2012ApJ...754...83B}.
\citet{2013ApJ...768L..37T} refined the $A_{\rm UV}-\beta$ relationship originally proposed by \citet{1999ApJ...521...64M}, introducing a Gaussian distribution for $\beta$ at each $M_{\rm UV}$ with a standard deviation $\sigma_{\beta} = 0.34$. Consequently, the average UV extinction $\langle A_{\rm UV} \rangle$ can be expressed as 
\begin{equation}
\langle A_{\rm UV} \rangle = C_0 + 0.2 C_1^2\ln{(10)} \sigma_{\beta}^2 + C_1\langle \beta (z,M_{{\rm UV}})\rangle,
\end{equation}
where $\sigma_{\beta} = 0.34$, $C_0=4.54$, and $C_1=2.07$ \citep{2011ApJ...726L...7O}. 
Besides, \citet{2015ApJ...813...21M} described $\langle \beta \rangle$ as
\begin{equation}
\begin{aligned}
&\langle \beta(z,M_{{\rm UV}})\rangle = \\
&\left\{ \begin{array}{ll}
 (\beta_{M_0}(z)-c)\exp{\bigg[ - \frac{\frac{{\rm d}\beta}{{\rm d} M_0}(z)[M_{{\rm UV}} -M_0]}{\beta_{M_0}(z)-c} \bigg] } +c, & M_{{\rm UV}} \ge M_0 \\
\frac{{\rm d}\beta}{{\rm d}M_0}(z)[M_{{\rm UV}} -M_0] + \beta_{M_0}(z), & M_{{\rm UV}}<M_0,
\end{array} \right .
\end{aligned}
\end{equation}
where $c=-2.33$, $M_0=-19.5$. The values for $\beta_{M_0}$ and ${\rm d}\beta/{\rm d}M_0$ were sourced from \citet{2014ApJ...793..115B}. This formulation allows for the transformation of observed UV LFs into an intrinsic framework by adjusting observed magnitudes, $M_{\rm UV}({\rm obs})$, using the relation $M_{\rm UV}({\rm int})=M_{\rm UV}({\rm obs})-\langle A_{\rm UV} \rangle (z,M_{\rm UV})$.

\subsection{Eliminating the artificial biases}

Subsequently, we attempt to mitigate the artificial biases present in the observations of the UV LFs. The UV LFs at a specific magnitude is calculated by dividing the number of detected galaxies by the comoving volume of the surveyed area. In our analysis, one objective is to derive various cosmological parameters from the UV LFs. Accordingly, it is essential to rescale the observations and their associated errors, e.g.,
\begin{equation}
\begin{aligned}
\Phi_{\rm UV}({\rm new})&=\Phi_{\rm UV}({\rm old}) \times \frac{V_{\rm c}({\rm old})}{V_{\rm c}({\rm new})}, \\
\sigma_{\rm UV}({\rm new}) &=\sigma_{\rm UV}({\rm old}) \times \frac{V_{\rm c}({\rm old})}{V_{\rm c}({\rm new})},
\end{aligned}
\end{equation}
as suggested by \citet{2022PhRvD.105d3518S}. However, this adjustment may prove to be inadequate. Considering a standard likelihood function 
\begin{equation}
\mathcal{L} = \prod^{N}_{i}\frac{1}{\sqrt{2\pi} \sigma_{{\rm UV},i}} \exp{\bigg[-\frac{1}{2}\bigg(\frac{\Phi_{\rm model}(M_{\rm i}) - \Phi_{{\rm UV},i}}{\sigma_{{\rm UV},i}} \bigg)^2 \bigg]},
\end{equation}
it is evident that a smaller $\sigma_{\rm UV}$ leads to an increased likelihood value. Consequently, the estimation of cosmological constants such as $H_0$ or $\Omega_{\rm m}$ could be significantly underestimated in Bayesian analysis due to the variability of $\sigma_{\rm UV}$.

To ensure that the ``observations" and their associated errors are insensitive to cosmological parameters and remain constant during Bayesian inference, we transform the  ``observations" from density $\Phi_{\rm UV}$ and absolute magnitude $M_{\rm UV}$, to the directly observed quantities: the number $N$ of galaxies and their apparent magnitude $m_{\rm UV}$, i.e,

\begin{equation}\label{eq:6}
\begin{aligned}
N & = V_{\rm c}({\rm old}) \Phi({\rm old}),\\
m_{\rm UV} & =M_{\rm UV}({\rm old}) + 5\log_{10} \frac{d_{\rm L}({\rm old})}{\rm Mpc} + 25,
\end{aligned}
\end{equation}
where the old values are based on the cosmological parameters that each UV LFs observation assumed.

This treatment also mitigates the bias across various galaxy surveys, which arises from differing assumptions in cosmological frameworks. Consequently, the UV LF model referenced in \autoref{eq:1} and \autoref{eq:2} would be modified accordingly.

With the above corrections, the complete conversion is given by
\begin{equation}
\begin{aligned}
N({\rm new})& =V_{\rm c}({\rm old}) \Phi_{{\rm UV}}({\rm old})\bigg|\frac{\Delta m_{\rm UV}({\rm old})}{\Delta m_{\rm UV}({\rm new})}\bigg|,\\
\sigma_{N}({\rm new})& =V_{\rm c}({\rm old}) \sigma_{{\rm UV}}({\rm old})\bigg|\frac{\Delta m_{\rm UV}({\rm old})}{\Delta m_{\rm UV}({\rm new})}\bigg|,\\
\end{aligned}
\end{equation}
where
\begin{equation}
\begin{aligned}
&m_{\rm UV}({\rm new})   = M_{{\rm UV}}({\rm old}) - \langle A_{{\rm UV}}\rangle \big(z,M_{{\rm UV}}({\rm old})\big)\\
& \qquad\qquad\qquad+5\log_{10}\frac{d_{{\rm L}}({\rm old})}{\rm Mpc}+25,\\
&\bigg|\frac{\Delta m_{\rm UV}({\rm old})}{\Delta m_{\rm UV}({\rm new})}\bigg| = \\
&\frac{\Delta M_{{\rm UV}} ({\rm old})}{\Delta M_{{\rm UV}} ({\rm old})-\langle A_{{\rm UV}} \rangle \big( z, M_{{\rm UV}}^{+}({\rm old})\big)+ \langle A_{{\rm UV}} \rangle \big( z, M_{{\rm UV}}^{-}({\rm old}) \big)},
\end{aligned}
\end{equation}
$\Delta M_{{\rm UV}}$ represents the bin width, and $M_{\rm UV}^{+/-}({\rm old})=M_{\rm UV}\pm\Delta M_{{\rm UV}}/2$. It should be noticed that the comoving volume $(V_{\rm c} \sim \frac{survey \, field}{all-sky \, filed} \times V_{\rm c}(total)$) originally represents the characteristic value that relies on the field of each galaxy surveys. Because both the model and the observations are multiplied by $V_{\rm c}$ in model fitting, the term of $\frac{survey \, field}{all \, sky \, filed}$ can be removed for 
logical simplification. Therefore, all of the $V_{\rm c}$ represent the value of the total-comoving volume in this work.

\subsection{Redshift-dependent star formation efficiency}
SFE is a critical ingredient in UV LF model, as it directly influences SFR which further impacts UV LFs. Previously, it was widely assumed that SFE was either constant or solely dependent on the mass of DM halos, i.e.,
\begin{equation}
\epsilon=\epsilon_0~{\rm or}~\frac{2\epsilon_{\rm N}}{\big( \frac{M_{\rm h}}{M_1} \big) ^{-\beta} + \big( \frac{M_{\rm h}}{M_1} \big)^{\gamma}},
\end{equation}
while being independent of redshift.

\begin{figure}[ht!]
	\centering
	\includegraphics[width=0.47\textwidth]{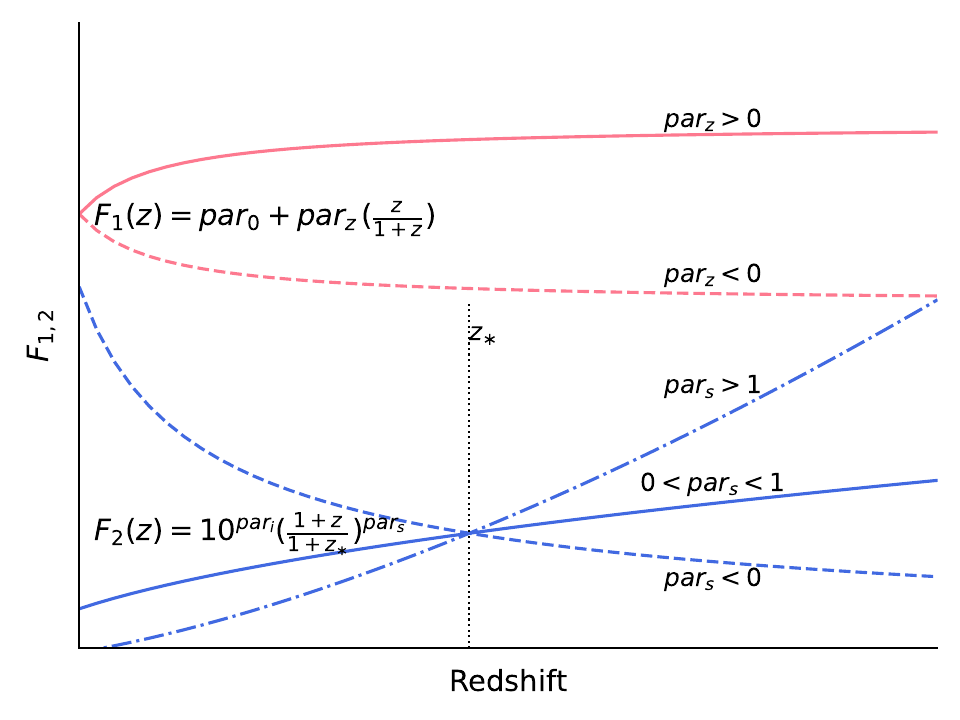}
	\caption{\small The variation of the parameter in SFE. The red ($F_1(z)$) and blue ($F_2(z)$) colors represent the models defined by \citet{2018MNRAS.477.1822M} and \citet{2022PhRvD.105d3518S}, respectively. For example, $\epsilon_{\rm N}$ is described by $\epsilon_0+\epsilon_z (\frac{z}{z+1})$ following $F_1(z)$ or $10^{\epsilon_i} (\frac{1+z}{1+z_*})^{\epsilon_s}$ following $F_2(z)$. Other parameters $M_1$, $\gamma$ and $\beta$ have the same definitions. Therefore, these two different formulae need 2 or 3 free variables and present different variable spaces. $z_*$ presents the specific redshift value defined in \autoref{eq:9}.}
	\label{fig:diagram}
\end{figure}

Recent studies by \citet{2018MNRAS.477.1822M} and \citet{2022PhRvD.105d3518S}, however, have proposed models where SFE varies with redshift, incorporating parameters such as $\epsilon_{\rm N}$, $M_1$, $\beta$, and $\gamma$ that evolve over time.
\autoref{fig:diagram} provides a comparative overview, illustrating how these parameters evolve with redshift according to two different definitions. Notably, due to significant variability observed in the blue lines of the diagram, we adopted definitions of $F_2(z)$ where both $\gamma$ and $\beta$ vary with redshift.
Consequently, the formulation of these SFE parameters is expressed as
\begin{equation}\label{eq:9}
par(z)=10^{par_i} \bigg(\frac{1+z}{1+z_{*}}\bigg)^{par_s},
\end{equation}
where $z_*$ is the specific redshift, and $par_i$, $z_{*}$ and $ par_s$ are the free variables for each SFE parameter.

\subsection{The effect of stochasticity}\label{sec:2.4}
Except for the aforementioned data corrections and model-dependent assumptions, the stochastic behavior in the SFR of high-redshift galaxies fluctuates the UV LFs and increases the uncertainties of parameter estimations. Such stochasticity can be driven by the process of bursty star formation and shows an increasing tendency at higher redshift, which has been used to explain the abundance of bright galaxies at $z\gtrsim10$ \citep{2023MNRAS.526.1324Q,2023ApJ...955L..35S,2024MNRAS.527.5929Y} that JWST detected recently. Following \citet{2023MNRAS.519..843M}, we add the stochasticity in accretion histories \citep{2018ApJ...868...92T} and consider a fraction of dex log-normal scatter ($\sigma_{\rm MAR}$) in DM halo accretion rate at fixed $M_{\rm h}$. In detail, the evolution of the stochasticity with redshift increasing is assumed to follow \autoref{eq:9} with $\sigma_{{\rm MAR}_s} > 0$, in a manner that aligns as analogical as possible with recent observations. Given that stochastic effects are more pronounced at higher redshifts, 
$\sigma_{\rm MAR}$ is assumed to be within $[-0.3, 0.3]$ ($[-0.1, 0.1]$) at $z\sim9$ ($z<8$). The corresponding variation in absolute magnitude is $\sigma_{M_{\rm UV}} \sim 0.75 \, (0.25)$, consistent with the variability obtained by \citet{2023MNRAS.525.3254S}, \citet{2024arXiv240504578K}, and \citet{2024ApJ...963...74W}.

\subsection{Bayesian analysis}\label{sec:2.5}
With the comprehensive UV LF model in hand, we aim to investigate the evolution of SFE and cosmological parameters. To accommodate the asymmetric and unequal error bars, we employed an asymmetric Gaussian distribution \citep{2013ApJ...778...66K} as the likelihood function over the redshift range from 4 to 10, i.e.,
\begin{equation}\label{eq:llh}
{\rm Likelihood} = \prod^{N}_{i} {{\rm AN}}(f(x_i)-y_i | c_i, d_i),
\end{equation}

where $(x_i, y_i)$ are the observational data points, $f(x_i)$ represents the predicted values, $d_i$ serves as a scale parameter, and $c_i$ is the asymmetry parameter. The priors for all primary parameters are detailed in \autoref{Tab:prior}.
For an optimal balance between accuracy and efficiency, we implemented the nested sampling method, utilizing {\tt Nessai} for Bayesian parameter estimation. We configured the analysis with 1000 live points\footnote{Preliminary tests with 500 live points yielded identical results. Given the recommendation to set the number of live points higher than the dimensionality of the parameter space \citep{2022NRvMP...2...39A}, our configuration ensures adequate convergence.} and an evidence tolerance of 0.5 to terminate the sampling process.

\begin{table}
	\begin{ruledtabular}
		\caption{Prior distributions of the parameters for SFE, stochasticity and cosmology}\label{Tab:prior}
		\begin{tabular}{lcc}
			Parameters  &Priors of parameter inference    \\ \hline            
			$\epsilon_{i}$       &Uniform(-3, 3)\\
	        $\epsilon_{s}$       &Uniform(-3, 3)\\
	        $M_{1,i}$            &Uniform(7, 15)\\
            $M_{1,s}$            &Uniform(-7, 7)\\
            $\beta_i$            &Uniform(-3, 3)\\
            $\beta_s$            &Uniform(-3, 3)\\
	        $\gamma_i$           &Uniform(-3, 3)\\
	        $\gamma_s$           &Uniform(-3, 3)\\
	        $\sigma_{{\rm MAR}_i}$           &Gaussian(0, 0.3)\\
	        $\gamma_{{\rm MAR}_s}$           &Uniform(0, 1)\\
	        $z_*$                &Uniform(0, 10)\\
	        $\sigma_8$  &0.812 or Uniform(0.1, 1.5)\textsuperscript{a}\\
		\end{tabular}
	\end{ruledtabular}
	\begin{tablenotes}
	 \item[a] \textsuperscript{a} Note that, in the first scenario (discussing the evolution of SFE and reionization process), we use the former; otherwise, we use the latter.
	\end{tablenotes}
\end{table}

\begin{figure*}[ht!]
	\centering
	\subfigure{
	    \hspace{-6mm}
		\includegraphics[width=0.49\textwidth]{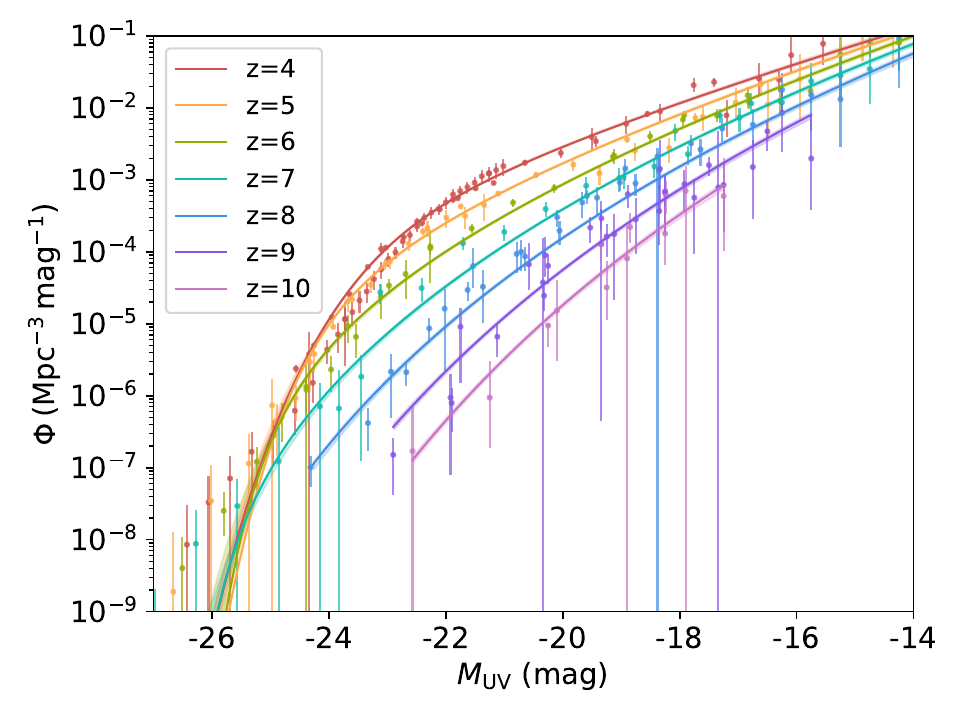}
		\hspace{2mm}
		\includegraphics[width=0.49\textwidth]{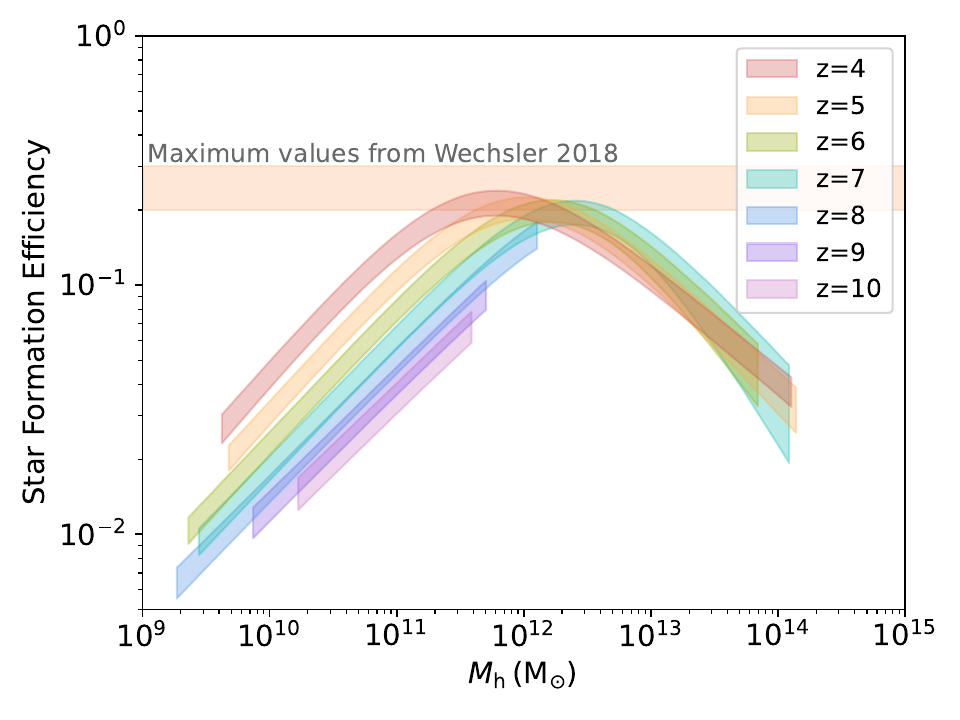}}
	\caption{\small The left panel shows the observations of UV LFs (extinction corrected) and the best-fit results in the redshift range of $4\le z \le 10$. The right panel shows the corresponding distributions of SFE. The colored regions are at $68\%$ credible level. The light orange region represents the range of maximum value (20\%-30\%) of SFE \citep{2018ARA&A..56..435W}. Due to the redshift-dependent nature of SFE, which connects all observations at each redshift, the estimated UV LFs are tightly constrained.}
	\label{fig:UV LF}
\end{figure*}

\begin{figure}[ht!]
	\centering
	\includegraphics[width=0.49\textwidth]{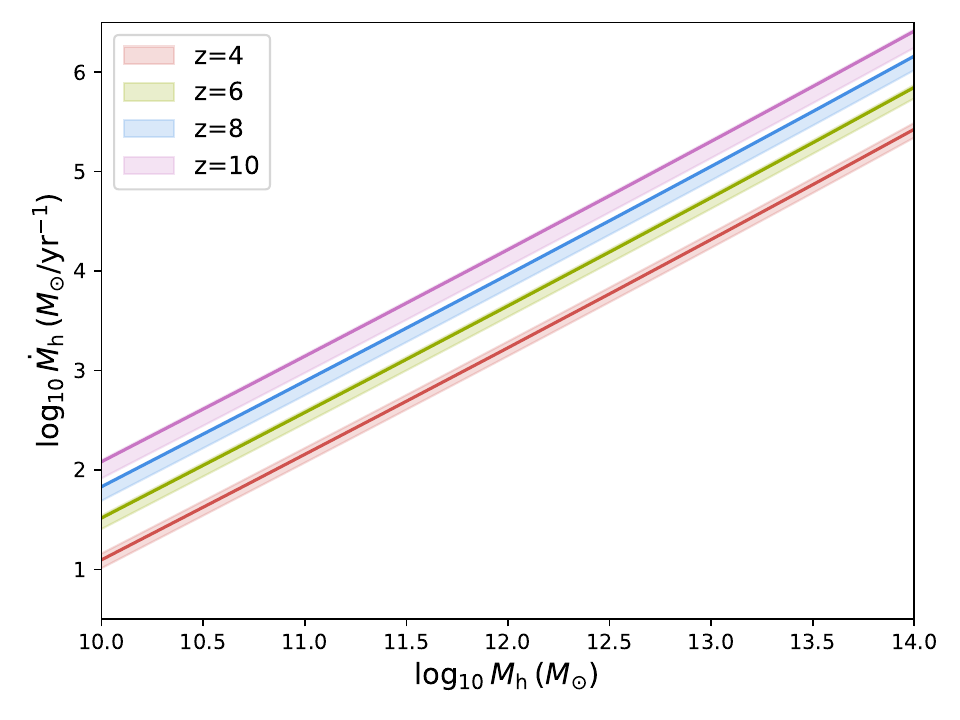}
	\caption{\small The posterior distributions of DM halo accretion rate at $z=4, \, 7$, and $10$. The solid lines represent the best-fit values. The colored regions are at $90\%$ credible level and represent the value of $\sigma_{\rm MAR}$.}
	\label{fig:accretion}
\end{figure}

\section{Results}
In this section, we present the fitting results under two scenarios. Initially, all parameters listed in \autoref{Tab:prior} are evaluated within $\rm \Lambda$CDM framework \citep{2020A&A...641A...6P}. This allows for an in-depth exploration of various reionization processes through the analysis of the derived UV luminosity density. Then, we assume $\sigma_8$ to be free and consider several additional datasets to evaluate the capability of constraining cosmological parameters by UV LFs. 

\subsection{The Evolution of SFE and UV LFs}\label{sec:3.1}
As shown in \autoref{fig:UV LF}, we present the best-fit and posterior results of UV LFs and SFE, alongside the observations. The corresponding posterior distributions are shown in \autoref{fig:pos_SFE}\footnote{It is worth noticing that we have tested the assumption that the SFE do not evolve with redshift ($par_s=0$). The (logarithmic) Bayes factor of the redshift-evolved SFE scenario compared to the non-evolving scenario is $\ln{\mathcal B}= 396.2$, which is strongly in favor of the redshift-evolution of the SFE.}. For a better view, we present the UV LFs rather than the numbers of galaxies that are used practically. It is evident that the UV LF model fits well with the observations except for the data points at the bright end ($M_{\rm UV}<-23 \rm mag$), which may be attributed to the assumption of dust extinction or the extra contribution of active galactic nuclear (AGN) in addition to the galaxy UV LFs. At the redshift range of $4\ge z\ge 7$, \citet{2022ApJS..259...20H} confirmed that the AGN LFs dominate at $M_{\rm UV}<-24 \, \rm mag$ and the galaxy LFs dominate at $M_{\rm UV}>-22 \, \rm mag$. They also found that the bright-end excess results in galaxy UV LFs is better fitted with the double-power-law function rather than the classical Schechter function. One possible explanation for the discrepancy is the empirical $\beta_{\rm UV} - M_{\rm UV}$ relation, which is not well-constrained at very bright magnitudes. Additionally, the true attenuation curve for high-redshift galaxies remains poorly understood \citep{2019MNRAS.487.1844M,2020A&A...643A...4F,2021MNRAS.502.3210L}. As a result, our model may have overestimated the dust obscuration, leading to larger uncertainties at the bright end. Furthermore, \citet{2022ApJS..259...20H} also suggested that even after subtracting the number density of AGNs based on spectroscopic galaxy fractions, hidden AGNs could still be present within the galaxy LFs, further contributing to the observed bright-end excess. Since our model relies on the shape of SFE, it is more analogous to the Schechter function and shows slight inconsistency with the observations at the bright end. Besides, a fraction of data shows better fitting extent with the empirical model at $z=4$, which arises from their smaller measuring uncertainties and the more harmonious tendencies towards higher-redshift results.

The corresponding distributions of the evolved SFE are consistent with other researches (such as Fig.~2 in \citet{2018ARA&A..56..435W}). The SFE peaks at $M_{1}\sim 10^{11.79}M_\odot$ for $z=4$ and $10^{12.84} \, \rm M_{\odot}$ for $z=10$ (note that the latter is only obtained for extrapolation since the extremely bright end of the UV LFs has not been detected at $z\ge8$).
At low halo masses, the profile of SFE slope is determined by $\beta$, which remains almost constant with $\beta \sim 0.50$ across wide redshift range. Similar conclusions are found in \citet{2024arXiv240702674F}, also. Conversely, at high masses, $\gamma$ dominates the profile which varies from $\sim 0.43$ to $\sim 1.34$, resulting in a notable steepening of the SFE. However, at $z\ge8$, due to the absence of UV LFs at the bright end, the SFE profile at $M_{1} \ge 10^{12} \, \rm M_{\odot}$ can not be well constrained. Consequently, it remains inconclusive whether there is a reliable evolution of the SFE at high masses. As an extra supplementary, \autoref{fig:accretion} illustrates the stochastic effect referred in \autoref{sec:2.4}. The posterior distributions of the DM halo accretion rate $\dot{M}_{\rm h}$ are consistent with the prior assumption, which is in line with the expectation. To further validate the reliability of our estimated results, we present a simplified comparison between our findings and those from GALLUMI. Detailed discussions are provided in \autoref{sec:app3}.

\begin{figure}[ht!]
	\centering
	\includegraphics[width=0.49\textwidth]{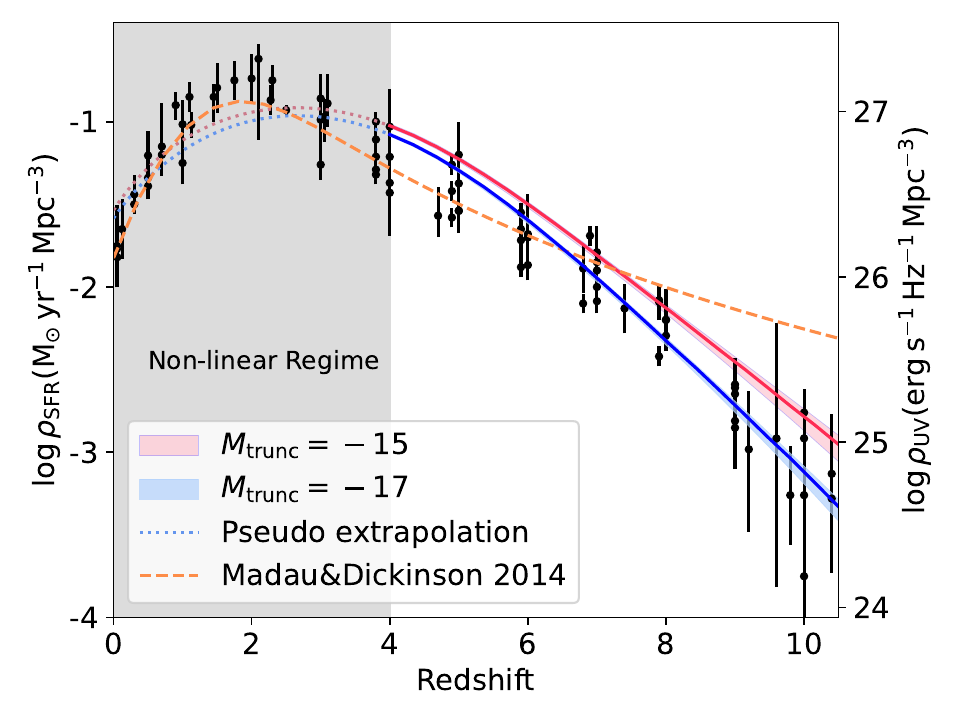}
	\caption{\small The evolution of the derived cosmic SFR density and UV luminosity density. The blue and red solid lines depict the best-fit values of UV LFs with $M_{\rm trunc} =-17 $ and $M_{\rm trunc} =-15$, respectively. The blue and red dotted lines are the pseudo extrapolation to low redshift range. The light blue and red shades indicate the 90\% credible regions. The orange dashed line denotes the analytical form derived by \citet{2014ARA&A..52..415M}. The grey region $0\le z \le 4$ implicates that the galaxy evolution is in a highly non-linear stage, which can not be analyzed using the UV LF model. The black error bars are not considered in Bayesian analysis and are solely included for comparison with our predictions. These observations encompass the radio, near-infrared, H$\alpha$ and UV bands at the redshift range of $4 \le z \le 10.5$ \citep{2014ARA&A..52..415M,2015ApJ...803...34B,2015ApJ...810...71F,2016MNRAS.459.3812M,2018MNRAS.477.1822M,2020ApJ...902..112B,2023ApJS..265....5H}.}
	\label{fig:rhoUV}
\end{figure}

Based on the fitting results of SFE, the entire UV LFs in the redshift range of $4\le z \le 10$ can be described. By integrating over a proper range of the absolute magnitude, the UV luminosity densities $\rho_{\rm}$ can be calculated by
\begin{equation}
\rho_{\rm UV} =\int^{M_{\rm trunc}}_{-\infty}\Phi_{{\rm UV}}(M)M_{{\rm UV}}(M){\rm d}M,
\end{equation}
where $M_{\rm trunc}$ denotes the truncation magnitude of the UV LFs. In general, $M_{\rm trunc} =-17 $ corresponds to the SFR of $0.3 \, \rm M_{\odot} \, yr^{-1}$ \citep{2015ApJ...810...71F,2018ApJ...855..105O,2020ApJ...902..112B,2022ApJS..259...20H}. Additionally, we consider another truncated limit with $M_{\rm trunc} =-15$ following \citet{2016MNRAS.459.3812M} and \citet{2018ApJ...854...73I}. As shown in \autoref{fig:rhoUV}, the UV luminosity densities derived from the fitted UV LFs are in accordance with other observations. The empirical equation Eq.(15) in \citet{2014ARA&A..52..415M} fits well with the observations at low redshifts (i.e., $z\leq 6.5$) but tends to overestimate the UV luminosity density and the SFR density at higher redshifts. 

\subsection{Properties of the Reionization}

\begin{table}
	\begin{ruledtabular}
		\caption{The observations of $x_{\rm H_I}$}\label{Tab:xHI}
		\begin{tabular}{lll}
redshift  &$x_{\rm H_I}$  &Publication  \\ \hline   
5.913   &$0.3_{-0.2}^{+0.2}$       &\citet{2014PASJ...66...63T}\\
6.6     &$0.15_{-0.15}^{+0.15}$    &\citet{2018PASJ...70S..13O}\\
6.6     &$0.08_{-0.05}^{+0.08}$    &\citet{2021ApJ...919..120M}\\
7.0     &$0.59_{-0.15}^{+0.11}$    &\citet{2018ApJ...856....2M}\\
7.0     &$0.7_{-0.23}^{+0.20}$     &\citet{2020ApJ...896...23W}\\
7.0     &$0.28_{-0.05}^{+0.05}$    &\citet{2021ApJ...919..120M}\\
7.09    &$0.48_{-0.26}^{+0.26}$    &\citet{2018ApJ...864..143D}\\
7.12    &$0.53_{-0.47}^{+0.18}$    &\citet{2024ApJ...971..124U}\\
7.3     &$0.55_{-0.25}^{+0.25}$    &\citet{2014ApJ...797...16K}\\
7.3     &$0.83_{-0.07}^{+0.06}$    &\citet{2021ApJ...919..120M}\\
7.44    &$0.65_{-0.34}^{+0.27}$    &\citet{2024ApJ...971..124U}\\
7.5     &$0.88_{-0.10}^{+0.05}$    &\citet{2019ApJ...878...12H}\\
7.5     &$0.21_{-0.19}^{+0.17}$    &\citet{2019MNRAS.484.5094G}\\
7.54    &$0.60_{-0.23}^{+0.20}$    &\citet{2018ApJ...864..143D}\\
7.6     &$0.49_{-0.19}^{+0.19}$    &\citet{2020ApJ...904..144J}\\
8.28    &$0.91_{-0.22}^{+0.09}$    &\citet{2024ApJ...971..124U}\\
9.91   &$0.92_{-0.10}^{+0.08}$  &\citet{2024ApJ...971..124U}\\
11.48   &$0.85_{-0.3}^{+0.15}$     &\citet{2023NatAs...7..622C}\\ \hline
5.6     &$<0.09$ &\citet{2015MNRAS.447..499M}\\
5.9     &$<0.06$ &\citet{2015MNRAS.447..499M}\\
6.3     &$<0.17$ &\citet{2006PASJ...58..485T}\\
7.0       &$<0.6$  &\citet{2015MNRAS.446..566M}\\
7.3     &$>0.28$ &\citet{2021ApJ...923..229G}\\
7.88    &$>0.45$ &\citet{2023ApJ...947L..24M}\\ 
8.0       &$>0.76$ &\citet{2019MNRAS.485.3947M}\\
10.17   &$>0.9$  &\citet{2024ApJ...973....8H}\\
10.6    &$<0.88$ &\citet{2023ApJ...949L..40B}\\ 
		\end{tabular}
	\end{ruledtabular}
\end{table}

Since the UV luminosity density covers the range of $4\le z \le 10$, it can be used to analyze the process of reionization. \citet{2013ApJ...768...71R} described the evolution of the ionized hydrogen fraction $\mathcal{Q}_{\rm H_{II}}$ as
\begin{equation}
\begin{aligned}
\dot{\mathcal{Q}}_{\rm H_{II}} & =\frac{\dot{n}_{\rm ion}}{\langle n_{\rm H} \rangle} - \frac{\mathcal{Q}_{\rm H_{II}}}{t_{\rm tec}}, \\
\dot{n}_{\rm ion} & \equiv \langle f_{\rm esc}\xi_{\rm ion} \rangle \rho_{\rm UV},\\
\langle n_{\rm H} \rangle &= \frac{X_{\rm p} {\rm \Omega_{\rm b} \rho_{\rm c}}}{m_{\rm H}}, \\
t_{{\rm rec}} & = \frac{1}{C_{{\rm H_{II}}} \alpha_{{\rm B}}(T) (1+Y_{{\rm p}}/4X_{{\rm p}}) \langle n_{{\rm H}} \rangle (1+z)^3},
\end{aligned}
\end{equation}
where $\langle n_{\rm H} \rangle$ is the mean hydrogen number density, $\dot{n}_{\rm ion}$ is the production rate of ionizing photons, and $t_{\rm tec}$ is the average recombination time in the IGM. Other parameters are defined as follows: the primordial mass fraction of hydrogen $X_{\rm p}=0.75$, the primordial helium abundance $Y_{\rm p}=1-X_{\rm p}$, the critical mass density $\rho_{c}=8.535 \times 10^{-30} \, \rm g \, cm^{-3}$, the mass of hydrogen atom $m_{\rm H}=1.66 \times 10^{-24} \, \rm g$, the coefficient for case B recombination $ \alpha_{{\rm B}}=2.6\times 10^{-13}\, {\rm cm^3\,s^{-1}}$ \citep{2016MNRAS.460..417S}, and the clumping factor $C_{\rm H_{II}} = 2.9 \times \big[\frac{(1+z)}{6}\big]^{-1.1}$. The integral product of the ionizing photon production efficiency $\xi_{\rm ion}$ and the escape fraction $f_{\rm esc}$ are regarded as variables during model fitting. The evolution of $\mathcal{Q}_{\rm H_{II}}$ can be depicted after considering two extra boundary conditions, 
\begin{equation}
\mathcal{Q}_{\rm H_{II}} (z_0) =1 \quad {\rm and } \quad  \mathcal{Q}_{\rm H_{II}} (z_{\rm max}) =0, 
\end{equation}
where $z_0$ and $z_{\rm max}$ are the redshift at the beginning and the end of the reionization, respectively.

\begin{table}
	\begin{ruledtabular}
		\caption{Prior distributions and posterior results of the parameters for $\mathcal{Q}_{\rm H_{II}}$}\label{Tab:Qprior}
		\begin{tabular}{lllc}
			Parameters  &Priors  &Posterior results\textsuperscript{a} \\ \hline     
			$z_0$       &Uniform(0,10)   &$5.3^{+0.8}_{-1.0}\,(5.3^{+0.8}_{-1.0})$\\
			$z_{\rm max}$ &Uniform(10,30) &$18.8^{+7.2}_{-6.0}\,(19.0^{+6.7}_{-5.8})$\\
			$\langle f_{\rm esc} \xi_{\rm ion} \rangle $\textsuperscript{b} &Uniform(0,1) $\times 10^{25.5}$ &$0.3^{+0.1}_{-0.1}\,(0.4^{+0.1}_{-0.1}) \times 10^{25.5} $\\
		\end{tabular}
	\end{ruledtabular}
	\begin{tablenotes}
	 \item[a] \textsuperscript{a} The values in these rows represent $\rho_{\rm UV}$ follows $M_{\rm trunc}=-17$. The uncertainties correspond to the $68\%$ confidence level. 
     \item[b] \textsuperscript{b} Assuming $\xi_{\rm ion} =10^{25.6} \, (10^{25.8})$ \citep{2023MNRAS.526.1657T,2024Natur.626..975A}, the values of $f_{\rm esc}$ is $\sim 20\%$.
    \end{tablenotes}
\end{table}

Whereafter, the evolution of $\mathcal{Q}_{\rm H_{II}}$ can be constrained by the observations of the neutral hydrogen fraction $x_{\rm H_I}$ ($x_{\rm H_I}=1-\mathcal{Q}_{\rm H_{II}}$). These additional data were obtained through various methods, including analysis of the ${\rm Ly}\alpha$ and ${\rm Ly}\beta$ forests of quasars, distributions of ${\rm Ly}\alpha$ equivalent width, adsorptions in ${\rm Ly}\alpha$ damping wings, fractions of ${\rm Ly}\alpha$ emitters, the afterglow spectrum of the gamma-ray burst, and the Gunn Peterson troughs. Using the similar screening methodology of UV LF observations, each of the data points of $x_{\rm H_I}$ at the same redshift has been confirmed without any redundancy. All of the observation results are summarized in \autoref{Tab:xHI}. 

\begin{figure}[ht!]
	\centering
	\includegraphics[width=0.49\textwidth]{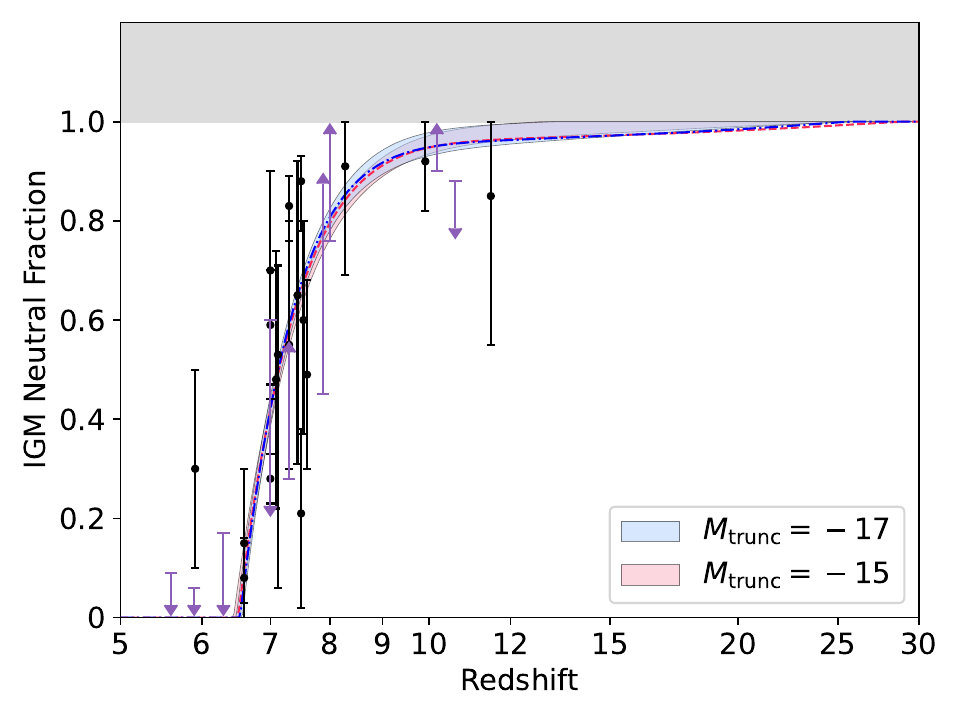}
	\caption{\small The evolution of the IGM neutral fraction. The black error bars and the purple bounds are the fitting data that listed in \autoref{Tab:xHI}. The red and blue lines correspond to the results shown in \autoref{fig:rhoUV}. The light-colored regions indicate the $90\%$ confidence level.}
	\label{fig:xHI}
\end{figure}

Besides, the Thomson optical depth to microwave background can be obtained by integrating $\mathcal{Q}_{\rm H_{II}}$,
\begin{equation}\label{eq:tau}
\tau_e(z)=\int^{z}_{0}\frac{c(1+z^{\prime})^2}{H(z')}\mathcal{Q}_{{\rm H_{II}}}\sigma_{{\rm T}}\langle n_{{\rm H}} \rangle \bigg( 1+\eta\frac{Y_{{\rm p}}}{4X_{{\rm p}}} \bigg){\rm d}z',
\end{equation}
where $\sigma_{\rm T}$ is the Thomson cross-section ($\sigma_{\rm T}=6.65\times10^{-29} \, {\rm m^2}$), $\eta=1$ at $z>4$ and $\eta=2$ at $z\le 4$. Thus, the observation of $\tau_{\rm e}$ ($\tau_{\rm e} = 0.054\pm0.007$ from \citet{2020A&A...641A...6P}) can be used to constrain $\mathcal{Q}_{\rm H_{II}}$. The Bayesian analysis follows the same process described in \autoref{sec:2.5}. For these upper and lower limits, the likelihood functions are assumed to be uniform within the limited range and to be a half-Gaussian beyond the boundary \citep{2017MNRAS.465.4838G}. The prior distributions and the posterior results of the free parameters are listed in \autoref{Tab:Qprior}. Using the $\rho_{\rm UV}$ derived from UV LFs (at $z>10$, the $\rho_{\rm UV}$ is extrapolated), the estimated evolution of $x_{\rm H_{I}}$ is shown in \autoref{fig:xHI}. Furthermore, we find out that the reionization started at $18.8^{+7.2}_{-6.0}\,(19.0^{+6.7}_{-5.8})$ and ended at $5.3^{+0.8}_{-1.0}\,(5.3^{+0.8}_{-1.0})$ when $M_{\rm trunc}$ is fixed to $-15\, (-17)$. Interestingly, such results are insensitive on the choice of $M_{\rm trunc}$. 

\begin{figure}[ht!]
	\centering
	\includegraphics[width=0.49\textwidth]{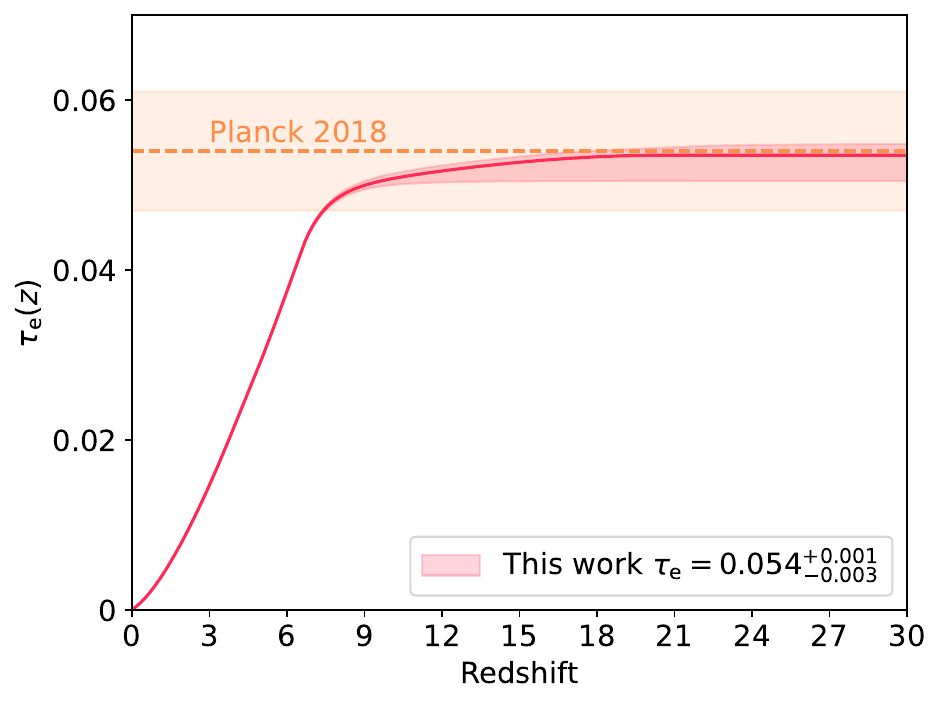}
	\caption{\small The predicted distribution of $\tau_{\rm e}$. The shaded pink region represents $68\%$ credible uncertainty range for $\tau_{\rm e}$. The light orange shade corresponds to the estimation provided by \citet{2020A&A...641A...6P}.}
	\label{fig:tau}
\end{figure}

Since \autoref{eq:tau} describes the relation between $\mathcal{Q}_{\rm H_{II}}$ and $\tau_{\rm e}$, we evaluate $\tau$ by fitting observations of $x_{\rm H_I}$ listed in \autoref{Tab:xHI}. Using $\rho_{\rm UV}$ with $M_{\rm trunc} =-15$, we present the extrapolated projections of $\tau_{\rm e}$ in \autoref{fig:tau}. Notably, our estimation is consistent with the result of \citet{2020A&A...641A...6P}, demonstrating the validity and reliability of our model.

\subsection{The constraints of the cosmological parameters}\label{sec:3.3}
As discussed in \autoref{sec:2}, the luminosity of galaxies relies on a specific cosmological framework, particularly $H_0$, $\Omega_{\rm m}$ and $\sigma_8$. However, these parameters degenerate with the parameters of SFE apparently, since SFE dictate SFR, $\sigma_8$ governs the relative density of DM halos, and $\Omega_{\rm m}$ and $H_0$ influence the density of galaxies through regulating $H(z)$. Therefore, additional constraints or supplementary dataset are required to narrow down the posterior parameter spaces within physical boundaries, especially for $H_0$ and $\Omega_{\rm m}$ because of their insensitivity to UV LFs. In light of this, we refer to the previous works of both \citet{2018MNRAS.477.1822M} and \citet{2022PhRvD.105d3518S}, considering three comparative cases:

$Case \,  1:$ To estimate $\sigma_8$, $\beta$ and $\gamma$ are assumed to be redshift-independence (i.e., $\beta_s=0$ and $\gamma_s=0$). 

$Case \, 2:$ To estimate $\sigma_8$, $\gamma$ is assumed to be redshift-independence (i.e., $\gamma_s=0$). 

$Case \, 3:$ To estimate $\sigma_8$, $\Omega_{\rm m}$ and $H_0$. Pantheon+ dataset \citep{2022ApJ...938..113S} and a prior constraint on $\Omega_{\rm b}$ \citep{2021JCAP...04..020P} are incorporated, and $\gamma$ is assumed to be redshift-independence (i.e., $\gamma_s=0$). The prior of the absolute B-band magnitude for the fiducial SN Ia is constrained within $[-20,-18]$ and the prior of $\Omega_{\rm b}$ follows the Gaussian distribution with $\mu=0.2233$ and $\sigma=0.00036$. The priors of $H_0$ and $\rm \Omega_{m}$ follow a Uniform distribution within $[50,90]$ and $[0.05,0.99]$, respectively.

Additionally, the maximum values of SFE are limited in $[0.01, 0.5]$,
which well covers the typical value $\sim 0.2-0.3$ found in previous analysis \citep{2018ARA&A..56..435W}. Furthermore, the parameter $M_{1,s}$ is bounded within $[-3,3]$, following the constraints from \citet{2022PhRvD.105d3518S}. 

\begin{figure}[ht!]
	\centering
	\includegraphics[width=0.49\textwidth]{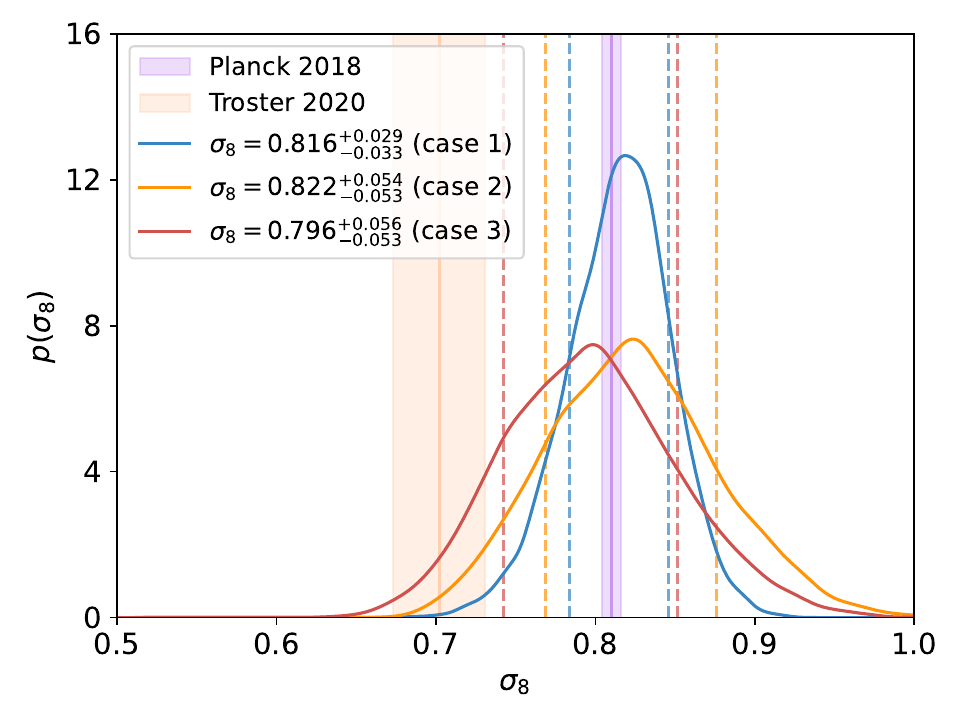}
	\caption{\small The results of $\sigma_8$ in three cases. To present the $\sigma_8$ tension, the purple and light orange regions represent the estimated $\sigma_8$ from \citet{2020A&A...641A...6P} and \citet{2020A&A...633L..10T}, respectively.}
	\label{fig:sigma8}
\end{figure}

\begin{figure}[ht!]
	\centering
	\includegraphics[width=0.45\textwidth]{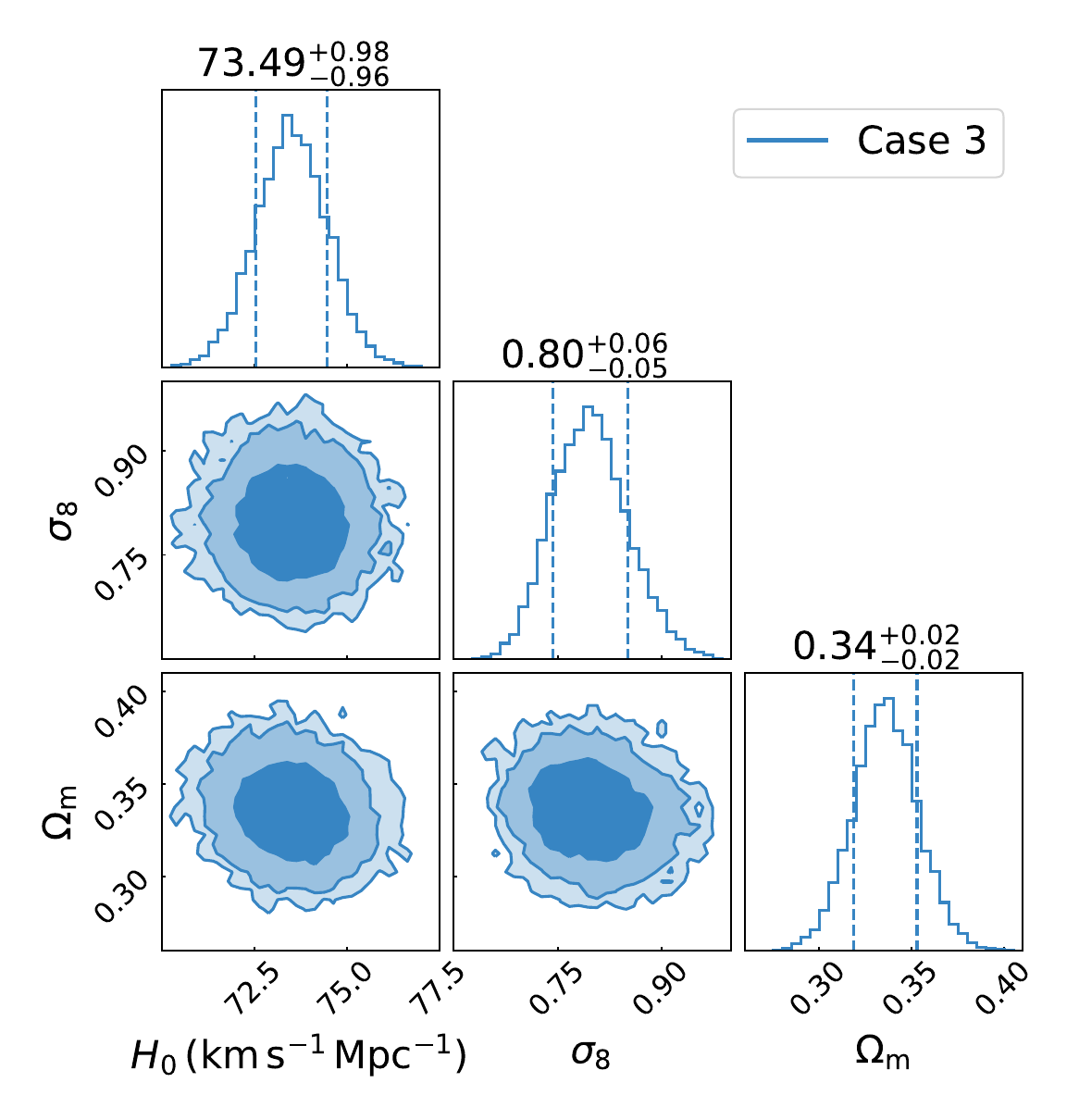}
	\caption{\small The posterior results of $H_0$, $\sigma_8$, and $\Omega_{\rm m}$ for $case \, 3$. The contours indicate the $68\%$, $95\%$, $99\%$ credible levels. The values are at $68\%$ credible level.}
	\label{fig:Omegamh0}
\end{figure}

The posterior distributions of estimated $\sigma_8$ are shown in \autoref{fig:sigma8}, which are more consistent with that obtained from early universe measurement. In $case\, 3$, the contours of $H_0$, $\Omega_{\rm m}$, and $\sigma_8$ are depicted in \autoref{fig:Omegamh0}, resembling the results of \citet{2022PhRvD.105d3518S}. Because we utilize a larger amount of data and align the observations within the same framework, our result of $\sigma_8 = 0.80^{+0.06}_{-0.05}$ reduces the uncertainty by $\sim 60\%$ compared to the estimation (i.e., $\sigma_8=0.76^{+0.12}_{-0.14}$) from \citet{2022PhRvD.105d3518S}. The complete posterior results of these three cases are shown in \autoref{fig:pos_case123}.  It is worth noting that all estimated parameters are well converged except for $M_{1,s}$, which tends to converge towards a higher value close to the upper limit as \citet{2022PhRvD.105d3518S} derived. Furthermore, due to the insensitivity to UV LF, the constraints on  $H_0$ and $\Omega_{\rm m}$ are not significantly improved in comparison to that derived by sole Pantheon+ dataset.

\section{Conclusions and discussions}
Over the passed several years, numerous/various galaxy surveys accumulated a wealth of observations. Especially, the observed UV LFs span a wide range of redshift, from the local universe to very high-redshift ($z\sim16$), providing the opportunity to study the epochs of cosmic dawn, reionization, and cosmic noon. In this work, we have developed a general UV LF model incorporating a redshift-dependent SFE and an alterable cosmological framework to explore the evolution of SFE, the process of reionization, and several cosmological parameters. By ‘‘correcting“ the observations to eliminate the effect of dust attenuation and reconcile the discrepancies of different cosmological frameworks, our results have higher precision compared with other works.

Under the framework of $\Lambda$CDM, the UV LFs within the redshift range of $z=4-10$ can constrain the evolution of SFE stringently. In \autoref{fig:UV LF}, the profile of SFE shows a clear tendency of evolution with redshift, particularly in the low mass range. The corresponding mass of the DM halos ($\sim 10^{12}\,\rm M_{\odot}$) at maximum SFE ($\sim 20\%$) presents a shift towards higher values, although further investigation is needed/warranted. Since the variable feedback mechanism and the environmental factors influence the relative strength of SFE, disentangling the contributions of each component solely from UV LFs poses challenges. Fortunately, dynamical simulations provide comparisons for these elements\citep[e.g.,][]{2022MNRAS.515.4929G,2024arXiv240407252S}, including the IMF, stellar radiation, stellar winds, supernovae, AGN and others. Therefore, once the profile of SFE can be constrained by other methods beyond UV LFs, the cosmological researches relying on UV LFs can yield more precise results, as the intrinsic degeneracy between SFE and cosmological parameters is alleviated. Furthermore, based on the derived UV luminosity density and abundant observations of the IGM neutral fraction, the beginning and ending redshifts of the reionization epoch are constrained to $18.8^{+7.2}_{-6.0}\,(19.0^{+6.7}_{-5.8})$ and  $5.3^{+0.8}_{-1.0}\,(5.3^{+0.8}_{-0.9})$ with $M_{\rm trunc}=-17$ and $-15$, respectively. If considering the observations of UV LFs at $z>10$, $\rho_{\rm UV}$ will be much higher than the extrapolated ones \citep{2023MNRAS.518.6011D,2023ApJS..265....5H}, but does not impact \autoref{fig:xHI} notably. We validate the reliability of our model by comparing the inferred Thomson scattering optical depth $\tau_{\rm e}$ with the result of \citet{2020A&A...641A...6P}. Nonetheless, our model is applicable only to the nonlinear regime and not to regimes with $z>10$, as uncertainties in stellar populations and dust attenuation lead to large uncertainties of SFE \citep{2022ApJ...938L..10I,2023ApJ...954L..48W} and other parameters.

On the other hand, we attempt to analyze cosmology by UV LF observations. Similar to previous works, we find that only introducing some reasonable and additional information, the cosmological parameters can be constrained within physically meaningful scopes. Within a specific parameter space, we obtain $\sigma_8=0.816^{+0.029}_{-0.033}$, which is consistent with the $\Lambda$CDM framework. Moreover, following \citet{2022PhRvD.105d3518S} and employing the same dataset and parameter spaces, the inferred $\sigma_8=0.796^{+0.056}_{-0.053}$ has a better precision, improved by $\sim 2-3$ times compared to the results reported by \citet{2022PhRvD.105d3518S}. It is in line with expectations since $\sigma_8$ is constrained by UV LFs mainly and the number of UV LF observations we used is over four times greater in number than they used, thereby reducing the error to $\sim 0.12/\sqrt{4}$. In addition to increasing the number of data points, expanding the survey area reduces the uncertainties associated with cosmic variance, thereby enhancing the accuracy of parameter estimation.

In the foreseeable future, various galaxy survey projects will make remarkable progresses. The Large Synoptic Survey Telescope \citep{2019ApJ...873..111I}, the Roman Space Telescope \citep{2021MNRAS.507.1746E}, Euclid \citep{2011arXiv1110.3193L}, the Extremely Large Telescope \citep{2007Msngr.127...11G}, and the China Space Station Telescope \citep{2011SSPMA..41.1441Z} will explore the Universe in deep field, providing extensive observational data with sufficient precision and covering a wide range of redshift. Furthermore, with the improving accuracy of dynamical simulation, the profile of SFE will be understood in depth, thereby breaking the intrinsic degeneracy between the SFE parameters and cosmological parameters. Consequently, through future observations of UV LFs and additional simulated constraints, there will be opportunities to analyze cosmological model, galaxy evolution, and the epoch of reionization with greater precision and to a more complete understanding in the future.

\begin{acknowledgements}
\section{Acknowledgements}
We thank the anonymous referee for helpful comments and suggestions. This work is supported in part by NSFC under grants of No. 11921003, No. 12233011 and 12303056; S.-P. T. acknowledges support from the General Fund (No. 2023M733736) of the China Postdoctoral Science Foundation and the Postdoctoral Fellowship Program of CPSF (GZB20230839); G.-W. Y. acknowledges support from the University of Trento and the Provincia Autonoma di Trento (PAT, Autonomous Province of Trento) through the UniTrento Internal Call for Research 2023 grant “Searching for Dark Energy off the beaten track” (DARKTRACK, grant agreement no.E63C22000500003, PI: Sunny Vagnozzi).

Software: 
{\tt Nessai} (\citet{2021PhRvD.103j3006W}), version 0.12.0, \url{https://nessai.readthedocs.io/en/latest/}, {\tt HMFcalc} (\citet{2014ascl.soft12006M}), \url{https://github.com/halomod/hmf/}
\end{acknowledgements}

\appendix
\section{The contributions of cosmic variance}
In this section, we present the impact of cosmic variance across different magnitudes and redshift. As an example, we list the observations from \citet{2021AJ....162...47B}. The total uncertainties used in our analysis include both the observed uncertainties and the contributions from cosmic variance.

\begin{table}[ht!]
\begin{ruledtabular}
\centering
\caption{The rest-frame UV LFs \citep{2021AJ....162...47B} and the corresponding cosmic variance (CV) at $z\sim4-9$}
\label{Tab:4}
\begin{tabular}{cccc|cccc}
Area &$M_{\rm UV}$ &$\Phi$ &CV &Area &$M_{\rm UV}$ &$\Phi$ &CV \\
($\rm{arcmin}^2$) & &($\rm{mag}^{-1} \rm{Mpc}^{-3}$) & &($\rm{arcmin}^2$) & &($\rm{mag}^{-1} \rm{Mpc}^{-3}$) & \\ \hline  
&-22.69 &$0.000005\pm0.000004$ &$0.2871$\textsuperscript{a} &&&&\\
&-22.19 &$0.000015\pm0.000009$ &$0.2871$ & &-22.19 &$0.000001\pm0.000002$ &$0.2079$\\
&-21.69 &$0.000144\pm0.000022$ &$0.2283$ & &-21.69 &$0.000041\pm0.000011$ &$0.1657$\\
&-21.19 &$0.000344\pm0.000038$ &$0.1847$ & &-21.19 &$0.000047\pm0.000015$ &$0.1403$\\
&-20.69 &$0.000698\pm0.000068$ &$0.1525$ & &-20.69 &$0.000198\pm0.000036$ &$0.1219$ \\
311.7 &-20.19 &$0.001624\pm0.000131$ &$0.1292$  &770.2 &-20.19 &$0.000283\pm0.000066$ &$0.1075$\\
$z\sim4$ &-19.69 &$0.002276\pm0.000199$ &$0.1109$ &$z\sim7$ &-19.69 &$0.000589\pm0.000126$ &$0.0958$\\
&-19.19 &$0.003056\pm0.000388$ &$0.0984$ & &-19.19 &$0.001172\pm0.000336$ &$0.0862$\\
&-18.69 &$0.004371\pm0.000689$ &$0.0882$ & &-18.69 &$0.001433\pm0.000419$ &$0.0785$\\
&-17.94 &$0.010160\pm0.000920$ &$0.0760$ & &-17.94 &$0.005760\pm0.001440$ &$0.0697$\\
&-16.94 &$0.027420\pm0.003440$ &$0.0638$  & &-16.94 &$0.008320\pm0.002900$ &$0.0612$\\
&-15.94 &$0.028820\pm0.008740$ &$0.0547$ &&&&\\ \hline
&-23.11 &$0.000001\pm0.000001$ &$0.2989$ &&&&\\
&-22.61 &$0.000004\pm0.000002$ &$0.2305$ &&&&\\
&-22.11 &$0.000028\pm0.000007$ &$0.1814$ & &-21.85 &$0.000003\pm0.000002$ &$0.1603$\\
&-21.61 &$0.000092\pm0.000013$ &$0.1456$ & &-21.35 &$0.000012\pm0.000004$ &$0.1395$\\
&-21.11 &$0.000262\pm0.000024$ &$0.1192$ & &-20.85 &$0.000041\pm0.000011$ &$0.1234$\\
770.2 &-20.61 &$0.000584\pm0.000044$ &$0.1006$ &988.5 &-20.10 &$0.000120\pm0.000040$ &$0.1047$\\
$z\sim5$ &-20.11 &$0.000879\pm0.000067$ &$0.0865$ &$z\sim8$ &-19.35 &$0.000657\pm0.000233$ &$0.0902$\\
&-19.61 &$0.001594\pm0.000156$ &$0.0763$  & &-18.60 &$0.001100\pm0.000340$ &$0.0799$\\
&-19.11 &$0.002159\pm0.000346$ &$0.0683$ & &-17.60 &$0.003020\pm0.001140$ &$0.0696$\\
&-18.36 &$0.004620\pm0.000520$ &$0.0587$ &&&&\\
&-17.36 &$0.008780\pm0.001540$ &$0.0496$ &&&&\\
&-16.36 &$0.025120\pm0.007340$ &$0.0425$ &&&&\\ \hline
&-22.52 &$0.000002\pm0.000002$ &$0.2423$ &&&&\\
&-22.02 &$0.000014\pm0.000005$ &$0.1892$  &&&&\\
&-21.52 &$0.000051\pm0.000011$ &$0.1511$ & &-21.92 &$0.000001\pm0.000001$ &$0.1457$\\
&-21.02 &$0.000169\pm0.000024$ &$0.1251$ & &-21.12 &$0.000007\pm0.000003$ &$0.1273$\\
770.2 &-20.52 &$0.000317\pm0.000041$ &$0.1063$  &918.2 &-20.32 &$0.000026\pm0.000009$ &$0.1124$\\
$z\sim6$ &-20.02 &$0.000724\pm0.000087$ &$0.0929$ &$z\sim9$ &-19.12 &$0.000187\pm0.000150$ &$0.0949$\\
&-19.52 &$0.001147\pm0.000157$ &$0.0824$  & &-17.92 &$0.000923\pm0.000501$ &$0.0814$\\
&-18.77 &$0.002820\pm0.000440$ &$0.0709$ &&&&\\
&-17.77 &$0.008360\pm0.001660$ &$0.0598$ &&&&\\
&-16.77 &$0.017100\pm0.005260$ &$0.0518$ &&&&\\ \hline

\end{tabular}
\end{ruledtabular}
\begin{tablenotes}
	 \item[a] \textsuperscript{a} Because this data point exceeds the limitation of {\tt GALCV}, the value of CV is assumed to be the latter one.
    \end{tablenotes}
\end{table}

\section{The posterior distributions of the parameters of SFE model}

Here, we present the whole posterior distributions of SFE parameters mentioned in \autoref{sec:3.1} and the detailed results for Case 1, Case 2, and Case 3 in \autoref{sec:3.3}.

\begin{figure*}[ht!]
	\centering
	\includegraphics[width=0.9\textwidth]{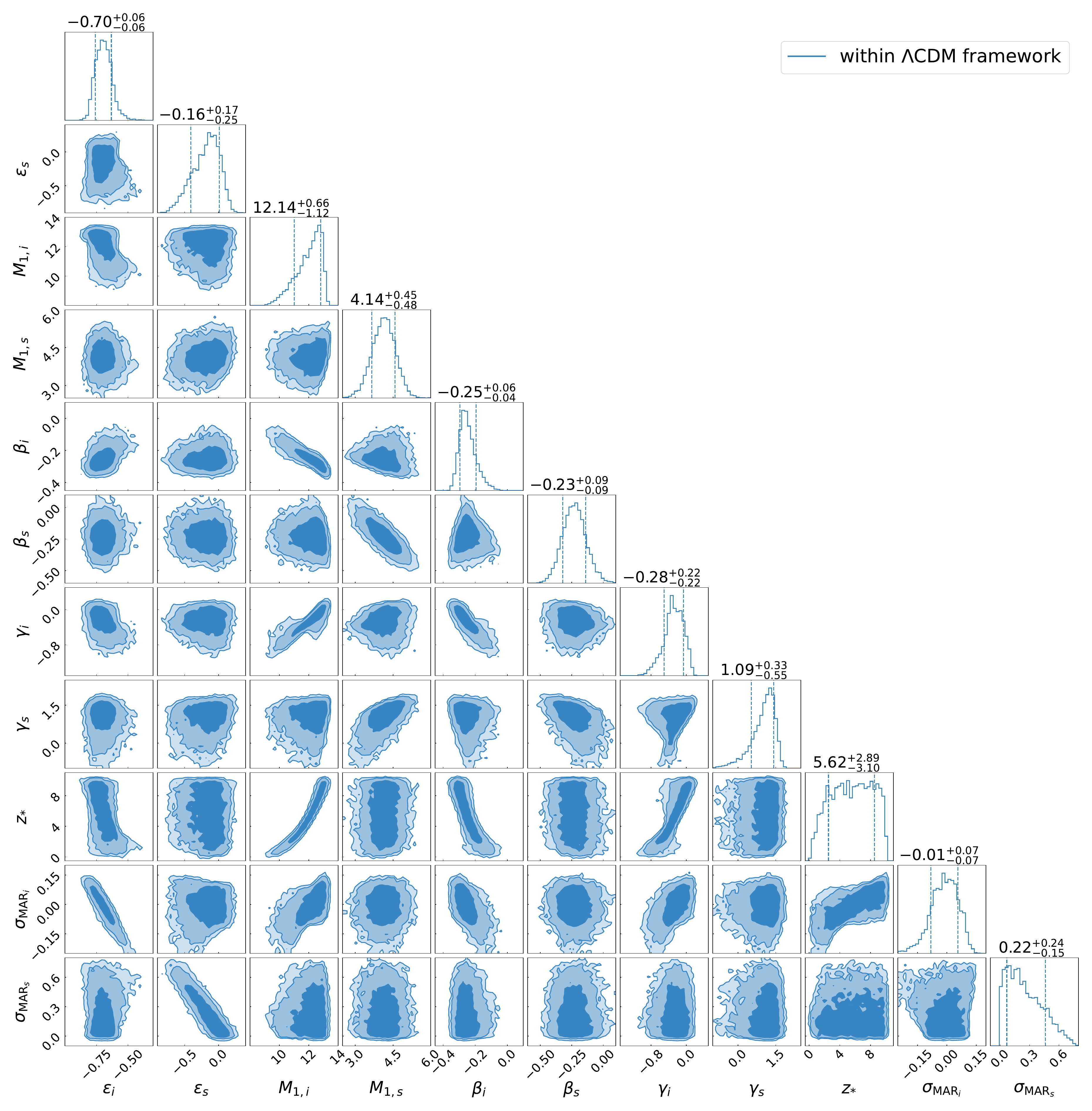}
	\caption{\small Posterior distributions of the SFE parameters for the UV LFs presented in \autoref{fig:UV LF}. The contours are at the 68\%, 95\%, and 99\% credible level, respectively. The reported values are at the 68\% credible level.}
	\label{fig:pos_SFE}
\end{figure*}

\begin{figure*}[ht!]
	\centering
	\includegraphics[width=0.99\textwidth]{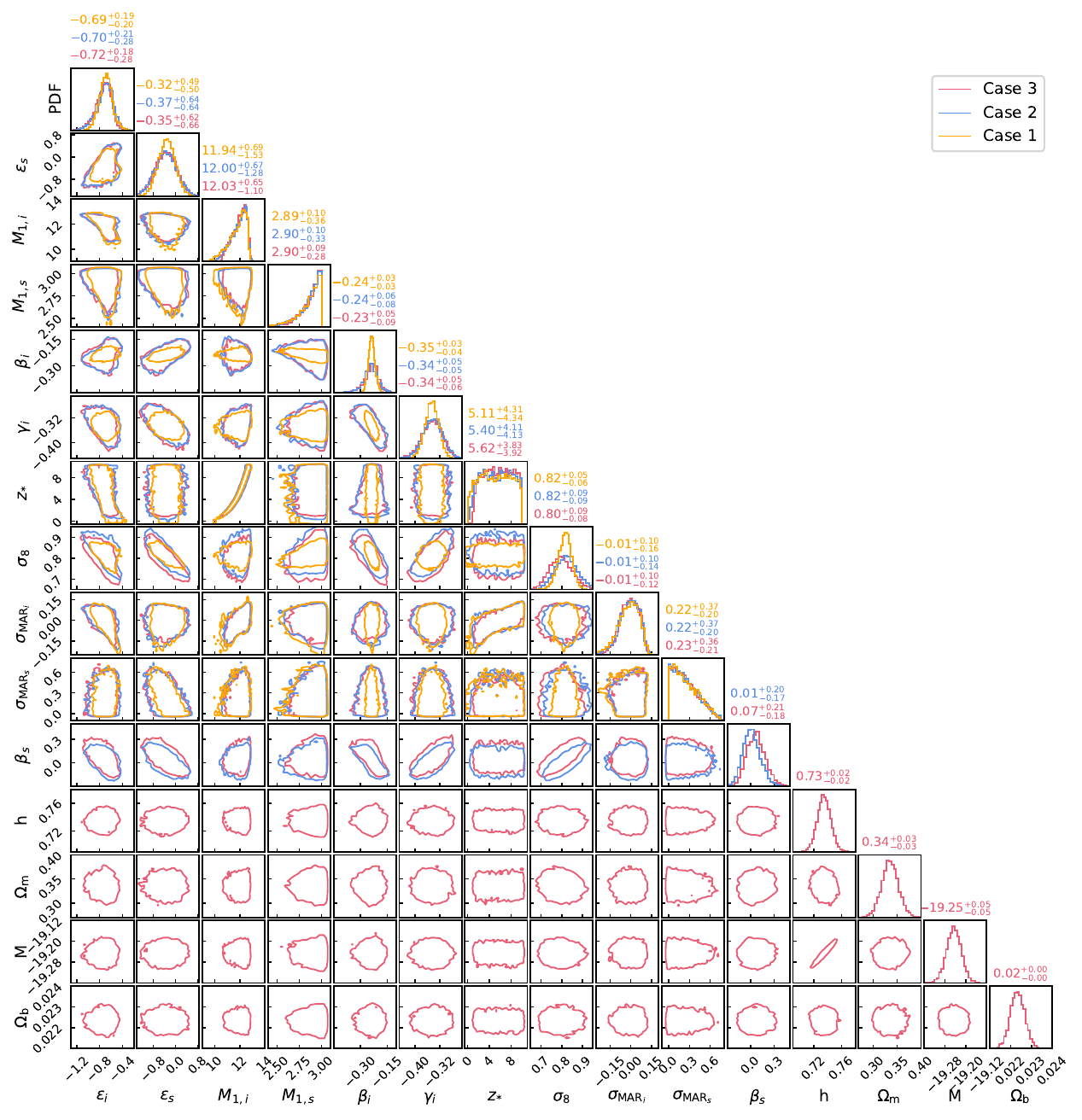}
	\caption{\small Complete posterior distributions of the SFE parameters and cosmological parameters for case 1, case 2 and case 3. All of the ranges are at the $90\%$ credible level.}
	\label{fig:pos_case123}
\end{figure*}

\section{The comparison between our results and GALLUMI}\label{sec:app3}
\begin{figure*}[ht!]
	\centering
	\includegraphics[width=0.8\textwidth]{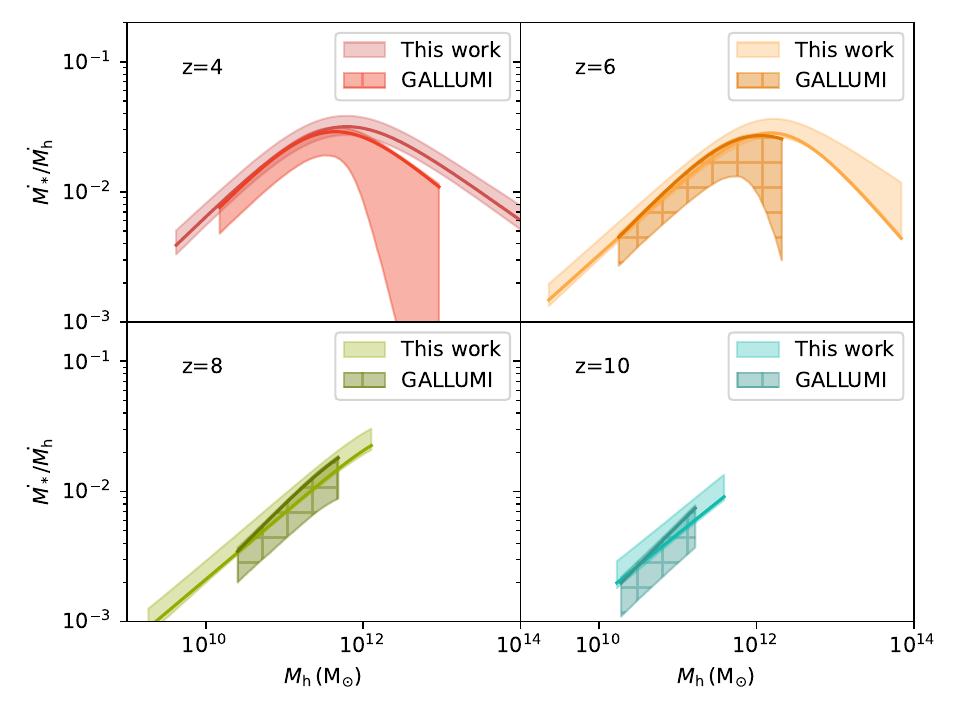}
	\caption{\small The posterior distribution of $\dot{M}_{*}/\dot{M}_{\rm h}$. The GALLUMI results use the same HST observations as \citet{2022PhRvD.105d3518S}. The solid lines indicate the best-fit values, while the colored regions represent the $90\%$ credible intervals, corresponding to the observed magnitude range.}
	\label{fig:MstarMh_compare}
\end{figure*}

In \autoref{fig:MstarMh_compare}, both the results are assumed to be within the $\Lambda CDM$ framework. Our findings provide tighter constraints on $\dot{M}_{*}/\dot{M}_{\rm h}$ due to the larger and more precise datasets used. The minor differences between the two results primarily stem from the variations in the observations. Additionally, a slight discrepancy in the definition of $\dot{M_{\rm h}}$ between the two models may also contribute to this mild discrepancy.

\clearpage
\bibliography{ref}
\bibliographystyle{aasjournal}
\end{document}